\documentclass[aps,pra,twocolumn,floatfix,nofootinbib]{revtex4-1}

\usepackage{xcolor}
\usepackage[utf8]{inputenc}  
\usepackage[T1]{fontenc}     %Output what you want e.g., é, ł, a, ü
\usepackage[british]{babel}  %Do hyphenation according to british english
\usepackage[sc,osf]{mathpazo}
\usepackage[scaled=0.86]{berasans}  % URL font that go well wtih palatino
\usepackage[colorlinks=true, citecolor=blue, urlcolor=blue]{hyperref}  %Hyperlinks (pink, green, blue)
\usepackage[babel]{microtype}  %Improves text justification
\usepackage{amsmath,amssymb,amsthm,bm,amsfonts,mathrsfs,bbm} %Usefull math packages

\usepackage{xspace}  %Useful to add space in macros
\usepackage{pgf,tikz}
\usepackage{xcolor}
\usepackage{multirow}
\usepackage{array}
\usepackage{bigstrut}
\usepackage{braket}
\usepackage{color}
\usepackage{natbib}
\usepackage{multirow}
\usepackage{mathtools}
\usepackage{float}
\usepackage[caption = false]{subfig}
\usepackage{xcolor,colortbl}
\usepackage{color}
\usepackage{physics}
\begin{document}

\title{Absolutely Secure Distributed Superdense Coding: Entanglement Requirement for Optimality}
\author{Sagnik Dutta$^1$}
\email{sagnikdutta17@gmail.com}
\author{Asmita Banerjee$^2$}
\email{asmibanerjee42@gmail.com}
\author{Prasanta K. Panigrahi$^1$}
\email{pprasanta@gmail.com}

\affiliation{$^1$Department of Physical Sciences, Indian Institute of Science Education and Research Kolkata ; Mohanpur, West Bengal 741246, India}
\affiliation{$^2$Narula Institute of Technology, Agarpara, Kolkata, West Bengal 700109, India}

\vspace{0.2 cm}
\begin{abstract}
    
    Superdense coding uses entanglement as a resource to communicate classical information securely through quantum channels. A superdense coding method is optimal when its capacity reaches Holevo bound. We show that for optimality, maximal entanglement is a necessity across the bipartition of Alice and Bob, but neither absolute nor genuine multipartite entanglement is required. Unlike the previous schemes, which can transmit either even or odd bits of information, we have demonstrated a generalized dense coding protocol using the genuine multipartite entangled GHZ state to send arbitrary information bits. Expressed in the eigenbasis of different Pauli operators, GHZ state is characterized by a unique parity pattern which enables us to formulate a security checking technique to ensure absolute security of the protocol. We show this method to be equally applicable in a scenario, where the resource information is distributed among spatially separated parties. Finally, optimizing the number of qubit(s) sent to Bob, we construct a distributed dense coding method, which completely depicts absolutely secure one way quantum communication between many to one party.
\end{abstract}
\maketitle

\section{Introduction}

    Quantum communication \cite{nielsen_chuang} is known to provide unparalleled level of security as opposed to its classical counterpart. It uses the unique features of quantum mechanics such as entanglement \cite{Ent,entanglement}, steering \cite{str}, quantum discord \cite{disc} and no cloning theorem \cite{Nocloning} to its advantage, which is not achievable through classical communication. Superdense coding \cite{Dense,Murli,Dense2,Dense3} and quantum teleportation \cite{Teleport,Teleport2,Teleport3,SJain} are the two most widely studied processes in this respect. These two share a close relationship in terms of using quantum resource like entanglement. While quantum teleportation transfers information between two parties using a distributed entangled state and a classical channel, superdense coding transfers classical information encoded in quantum mechanical bits using a distributed entangled state and a quantum channel.
    
    In the last two decades, Quantum Secure Direct Communication (QSDC) \cite{secret,secret2,secret3,secret4,secret5} has emerged as one of the important branches of quantum cryptography \cite{crypt,crypt2,crypt3}. Superdense coding is one such primary protocol to achieve QSDC because it provides security against eavesdropping. It uses maximally entangled pair of qubits in which the sender operates and sends just one qubit to transfer two bits of classical information. After Bennett and Wiesner \cite{Dense} demonstrated the superdense coding protocol to send two bits of information using EPR pair, various multipartite entangled states like Greenberger–Horne–Zeilinger (GHZ) state \cite{GHZ,GHZ2}, W states \cite{W}, cluster state \cite{Cluster}, Brown state \cite{Murli} was shown to successfully accomplish it. Generalized superdense coding has been done to send even number of bits $(2N)$ of information using N EPR pairs \cite{2N}. Saha and Panigrahi \cite{2N+1} proposed a generalized protocol for odd number of bits $(2N+1)$ of information using a GHZ states and $(N-1)$ EPR pairs.  
    
    Here, entanglement is used as a resource \cite{resource} which provides the quantum advantage in terms of qubits sent for sending the classical information as well as the security of this communication protocol. It is then required to inquire about the precise requirement of entanglement in the process. Since entanglement is the main resource behind all these communication protocols, detailed and extensive study of the same is an active research field \cite{resource2,Triangle}. Till now, entanglement classes upto four qubits \cite{three,four,four2,four3,three2} are completely documented. As the number of qubits increases, degrees of freedom for local quantum subsystems increase as well \cite{multi,multi2,multi3,multi4} and that makes our understanding of multipartite entanglement limited. Absolute maximally entangled (AME) \cite{AME,AME2,AME3} states are those multipartite states that have maximal entanglement across all bipartitions. When a multipartite state is not necessarily maximally entangled but non-separable across all bipartitions, one considers them as genuine multipartite entangled (GME) states. The requirement of these AME or GME states are yet to be analized in the context of superdense coding, which may also be achievable using mixed entangled states. However, that will never be optimal \cite{Mixed} because a N partite mixed state can be considered as a reduced state of a (N+1) partite pure entangled state. The other scenario of superdense coding, where the information is distributed among different senders and they want to send it to one particular receiver is known as the distributed superdense coding. It is known that the information is possible to send to the receiver only by applying local operations, without reducing the channel capacity \cite{DSC,DSC2,DSC3}. However, there does not exist any efficient method yet to do so.

    In Section \ref{Ent. req.}, we prove that optimized superdense coding using two qubits can take place only on AME states. For more than three qubit systems, neither AME nor GME states are required for optimal dense coding, but maximal entanglement across the bipartition of Alice and Bob is a necessity. Section \ref{N-separable} briefly describes the superdense coding protocol for even bits of information using a N-separable state. In Section \ref{Generalized}, we demonstrate a complete generalized version of superdense coding using the genuine multipartite entangled (GME) \cite{GME,GME2} N qubit GHZ state. Using the particular pattern present in GHZ state while written in different eigenbasis of Pauli operators we ensure absolute security of this protocol for ideal conditions in Section \ref{Security}. Using this present scheme we formulate a perfect distributed dense coding technique in Section \ref{D}. We also show that the capacity of all these protocols reache Holevo bound, making them optimal.
    
\section{Entanglement requirment for generalized Superdense coding}\label{Ent. req.}
    
    A pure state consisting n qudits of dimension d, $\ket{\psi}\in \mathbb{C}^{\otimes n}_d$ can be considered as a AME state iff every bipartition of the system such as A and B with $m=|B|\leq|A|=n-m$, is strictly maximally entangled i.e., $S(\rho_B)=m\log_2d$. As a consequence, for each bipartition the reduced density matrix of subsystem B must be of the form:
    \begin{equation}
        \rho_B=Tr_A\ketbra{\psi}=\frac{I_{d^m}}{d^m}, ~0<m\leq\frac{n}{2}
    \end{equation}
    Till now, AME states are quite limited in number. There does not exist any AME state for four, seven or more qubits. Similar to AME, a pure GME state is a pure state, for which each bipartition of the system such as A and B with $m=|B|\leq|A|=n-m$, the reduced state of both parties are mixed states i.e., $S(\rho_A),S(\rho_B)>0~\forall~0<m\leq\frac{n}{2}$.
    
    While sending N bit information through superdense coding, Alice operates a combination of four possible operations $\{I,\sigma_X,\sigma_Z,i\sigma_Y\}$ on her qubits to encode the message. That generates $2^N$ mutually orthogonal states which Bob measures and identifies the message. Suppose, Alice possess $|A|$ number of qubits, then maximum number of different states she can prepare is $4^{|A|}$. For the protocol to be `optimal' i.e., to reach its maximal capacity it must satisfy:
    \begin{equation}
        4^{|A|}\geq 2^N \implies |A|\geq \frac{N}{2}
        \label{cond1}
    \end{equation}
    Hence, to show entanglement requirement, we only need to focus on Bob's reduced subsystem.
    
    The capacity ($X$) of any particular superdense coding protocol \cite{Capacity,Capacity1} using given shared state ($\rho_{AB}$) is given as: 
    \begin{equation}
        X(\rho_{AB})=\log_2d_A+S(\rho_B)-S(\rho_{AB}),
        \label{capacity}
    \end{equation}
    where $d_A$ is the dimension of Alice's subsystem and $\rho_B$ is the reduced state of Bob. A superdence coding protocol is considered `optimal' when its capacity is maximum i.e., it reaches the Holevo bound. Therefore, to optimize the protocol we must use such a state that maximizes $X$ hence $S(\rho_B)$. From Eq. \ref{capacity}, it is evident that no mixed state can be used to get an optimized scheme because $S(\rho_{AB})$ is minimum only for pure states. Suppose Bob possesses $|B|$ qubits then from Eq. \ref{cond1}, $|B|\leq\frac{N}{2}\leq |A|$. To maximize $S(\rho_{B})$, the reduced density matrix of Bob (dimension $d_B$) must yield maximal entropy. That can only happen when, 
    \begin{equation}
        \rho_B=\frac{I_{d_B}}{d_B},~ 2\leq d_B=2^{|B|}\leq 2^{\frac{N}{2}}.
        \label{cond2}
    \end{equation}
    For any bipartite two qubit system Bob can only have one qubit with the marginal state $\frac{I}{2}$. Using Schmidt decomposition, any two qubit pure state $\chi$ can be written as $\chi=\sum_{i=0}^1\sqrt{\lambda_i}\ket{a_i}\ket{b_i}$, where $\{\ket{a_i}\}$ and $\{\ket{b_i}\}$ are sets of orthogonal basis vectors for the subsystem of Alice and Bob respectively. This Schmidt decomposed form of $\chi$ also yields the reduced state of Alice and Bob as $\rho_A=\sum_{i=0}^1\lambda_i\ket{a_i}\bra{a_i}$ and $\rho_B=\sum_{i=0}^1\lambda_i\ket{b_i}\bra{b_i}$. For maximally mixed $\rho_B$, we get $\lambda_0=\lambda_1=\frac{1}{2}$ for any chosen basis $\{\ket{b_i}\}$. Hence, the reduced state of Alice becomes $\rho_A=\frac{1}{2}(\ketbra{a_0}+\ketbra{a_1})$, which is exactly $I/2$ for the chosen basis $\{\ket{a_i}\}$ for Alice. So, the state $\chi$ must be a maximally entangled state. It implies that AME condition is a necessity to apply superdense coding using two qubits.
    
    As we increase the number of qubits used in the dense coding protocol, entanglement requirement changes. From condition \ref{cond2}, we can conclude that across the bipartition of Alice and Bob, maximal entanglement is necessary for optimized superdense coding to happen. We note that Eq. \ref{cond2} does not impose any condition across other bipartitions. It is not required for a state to be AME or even GME state to perform this protocol successfully. Any biseparable, triseparable or even N-separable state can be used, ensuring that across Alice and Bob's bipartition, the state is maximally entangled.
    
    If one view this dense coding scheme from security perspective, one obtains entanglement criteria for Alice's subsystem. For the protocol to be absolutely secure the marginal state of Alice must be a GME state. If it is not the same then we can write $\rho_A$ as:
    \begin{equation}
        \rho_A=\sum_{i}p_i\begin{pmatrix}
            a_{11} & \dots & a_{1m} \\
            \vdots & \ddots & \vdots\\
            a_{m1} & \dots  & a_{mm}
        \end{pmatrix}_{m\times m}\otimes
        \begin{pmatrix}
            b_{11} & \dots & b_{1n} \\
            \vdots & \ddots & \vdots \\
            b_{n1} & \dots  & b_{nn}
        \end{pmatrix}_{n \times n};
        \label{separable}
    \end{equation}
    with $m+n=|A|$. This implies that the state is separable across those m qubits versus rest bipartition. While sending those m qubits to Bob, Eve can manage to have access to all of those. Whatever the decoding scheme of Bob be for those m qubits, Eve can always apply the same and get the classical information encoded within those m qubits. So, the protocol will not remain completely secure.
    
\section{Generalized dense coding using N-separable states}\label{N-separable}
    In this section, we briefly outline a generalized superdense coding protocol to transfer even $(2N)$ bits of information. For this scheme, first proposed by Rigolin et. al. \cite{2N}, following N-separable $2N$ qubit state is used:
    \begin{equation}
        \ket{\psi_{2N}}=\ket{\phi^+}^{\otimes N}_{AB}; ~~ \ket{\phi^+}=\frac{1}{\sqrt{2}}\left( \ket{00}+\ket{11}\right)
    \end{equation}
    The first qubit of each $\ket{\phi^+}$ of this product state of N Bell states belongs to Alice and the other one to Bob. Alice encodes exactly two bits of classical information on each of her qubits by applying $\{I,\sigma_X,\sigma_Z,i\sigma_Y\}$ for the bit strings $\{00,01,10,11\}$ respectively and sends her qubits to Bob. Finally, Bob makes a Bell measurement on each of the N pairs. Thus, by sending only N qubits to Bob, she manages to transfer $2N$ bits of information. 
    
    Here, $\ket{\phi^+}$ being a maximally entangled state, both particles have a reduced state $\frac{I}{2}$. That makes the marginal state of the subsystems of both Alice and Bob $\rho_A=\rho_B=\frac{I}{2}^{\otimes N}=\frac{I_N}{2^N}$, which ensures the presence of maximal entanglement between Alice and Bob despite $\ket{\psi_{2N}}$ being a N-separable state. It also makes sure that the protocol remains safe from eavesdropping. Each qubit sent by Alice, being maximally mixed Eve can not extract any information from them, even if she has access to all of them.
    
\section{Generalized dense coding using genuinely entangled GHZ states }\label{Generalized}
    Previously, we illustrated an example of a superdense coding protocol to communicate even bit of information through an N-separable quantum channel. In this section, we provide a completely generalized new scheme of dense coding, showing that GME states like generalized GHZ states can be safely used as a dense coding channel with absolute security. First, we show the protocol using three qubit GHZ state and then based on that scheme we generalize it.
    
    \subsection{Three qubit GHZ state protocol}
    Since entanglement is necessary for efficient communication of any information, we consider the maximally entangled tripartite GHZ state $\ket{\psi_{3}}=1/\sqrt{2}\left( \ket{000}+\ket{111}\right)_{A_1A_2B}$, distributed among the two parties Alice and Bob. Here, the first two qubits belongs to Alice (denoted by subscript $A_1$ and $A_2$) and one to Bob (denoted by subscript B). Alice now applies suitable unitary transformations on her qubits, according to the classical information she wants to send. She chooses her unitary operations from the set $\{I,\sigma_X,\sigma_Z,i\sigma_Y\}$ and generates $16$ states out of which $8$ states will be mutually orthogonal. After Alice sends her qubits to Bob, he then performs a tripartite measurement in that GHZ basis to distinguish them uniquely and get the encoded information. The list of all unitary transformations Alice does to encode the information message and the respective states obtained are demonstrated in the following table:
    \begin{table}[htbp]
        \centering
        \begin{tabular}{|c|c|c|c|}
        \hline
            \textbf{Message} & \textbf{Unitary} & \textbf{Alternate} & \textbf{State} \\
            & \textbf{Operation} & \textbf{Operation} & \\
            \hline
            $000$ & $I\otimes I$ & $\sigma_Z\otimes \sigma_Z$ & $\ket{000}+\ket{111}$ \\
            $001$ & $I\otimes \sigma_X$ & $\sigma_Z\otimes i\sigma_Y$ & $\ket{010}+\ket{101}$ \\
            $010$ & $\sigma_X\otimes I$ & $i\sigma_Y\otimes \sigma_Z$ & $\ket{100}+\ket{011}$ \\
            $011$ & $\sigma_X\otimes \sigma_X$ & $i\sigma_Y\otimes i\sigma_Y$ & $\ket{110}+\ket{001}$ \\
            $100$ & $\sigma_Z\otimes I$ & $I\otimes \sigma_Z$ & $\ket{000}-\ket{111}$ \\
            $101$ & $\sigma_Z\otimes \sigma_X$ & $I\otimes i\sigma_Y$ & $\ket{010}-\ket{101}$ \\
            $110$ & $i\sigma_Y\otimes I$ & $\sigma_X\otimes \sigma_Z$ & $\ket{100}-\ket{011}$ \\
            $111$ & $i\sigma_Y\otimes \sigma_X$ & $\sigma_X\otimes i\sigma_Y$ & $\ket{110}-\ket{001}$ \\
            \hline
        \end{tabular}
        \caption{State obtained by Alice after applying suitable unitary operation (The coefficient $\frac{1}{\sqrt{2}}$ along with any global phase is removed for convenience)}
    \end{table}
    Here, all the eight states generated through Alice's operation are orthogonal to each other and hence forms the measurement basis for Bob. From the above table one can also notice that applying $\sigma_Z\otimes \sigma_Z$ on the first two qubits of $\ket{\psi_{3}}$ is same as applying $I\otimes I$. Again, application of $\sigma_Z$ switches the elements of $\{I,\sigma_Z\}$ and $\{\sigma_X,i\sigma_Y\}$ within themselves. Without loss of generality, let Alice choose her unitary operation for her first qubit from the set $\{I,\sigma_X,\sigma_Z,i\sigma_Y\}$ unbiasedly without any restriction. Then to obtain all unique states and to avoid the repetition of obtaining same state from different operations, she is left with only two choices for her second qubit. Those must be any one from $\{I,\sigma_Z\}$ and any one from $\{\sigma_X,i\sigma_Y\}$. Here, we choose them to be $I$ and $\sigma_X$.
    
    \subsection{Generalized Protocol}\label{GME}
    Using the previous technique applied for three qubit superdense coding, we generalize the concept for N qubits. For that we consider generalized N qubit GHZ state given as:
    \begin{equation}
        \ket{\psi_N}=\frac{1}{\sqrt{2}}\left(\ket{0}^{\otimes N}+\ket{1}^{\otimes N}\right)_{A_1\cdots A_{N-1}B}~\forall~N\in \mathbb{N} - \{1\}
    \end{equation}
    where first $N-1$ qubits belong to Alice and the last one qubit to Bob. Similarly, the unitary operation choice for three qubit protocol, here Alice chooses the unitary operation for first qubit from $\{I,\sigma_X,\sigma_Z,i\sigma_Y\}$ unbiasedly and for the rest of her qubits, she is left to choose any one from $\{I,\sigma_Z\}$ and also any one from $\{\sigma_X,i\sigma_Y\}$. Hence, we can write Alice's operation choice as:
    \begin{equation*}
        \mathcal{P}\otimes\mathcal{Q}^{\otimes N-2}~|~\mathcal{P}\in\{I,\sigma_X,\sigma_Z,i\sigma_Y\}, \mathcal{Q}\in \{I/\sigma_Z,\sigma_X/i\sigma_Y\}
    \end{equation*}
    To avoid the repetition of obtaining same state from different operations, we let Alice to have $4$ choices for first qubit, whereas for the rest of $N-2$ qubits, Alice has only $2$ choices each. So, total available unitary operations for Alice is $4\times2^{N-2}=2^N$ and by applying them on first $N-1$ qubits of $\ket{\psi_N}$, Alice will get $2^N$ different states. Now, superdense coding is possible only if these $2^N$ states become mutually orthonormal to each other. To prove orthonormality condition, we consider two general unitary operators $\mathcal{M}=\mathcal{P^{\prime}}\otimes\mathcal{Q^{\prime}}_{k}^{\otimes^{N-2}_{k=1}}$ and $\mathcal{N}=\mathcal{P^{\prime\prime}}\otimes\mathcal{Q^{\prime\prime}}_{k}^{\otimes^{N-2}_{k=1}}$ acting on $\ket{\psi_N}$. Taking inner product of these two states we get:
        \begin{align}
        &\left[\mathcal{M}\ket{\psi_N}\right]^{\dagger}\left[\mathcal{N}\ket{\psi_N}\right] \nonumber \\
        &=\left[\mathcal{P^{\prime}}\otimes\mathcal{Q^{\prime}}_{k}^{\otimes^{N-2}_{k=1}}~\frac{1}{\sqrt{2}}\left(\ket{0}^{\otimes N}+\ket{1}^{\otimes N}\right)\right]^{\dagger}\nonumber\\ &~~~~~~~~~~~~~~~~~~~~\left[\mathcal{P^{\prime\prime}}\otimes\mathcal{Q^{\prime\prime}}_{k}^{\otimes^{N-2}_{k=1}}~\frac{1}{\sqrt{2}}\left(\ket{0}^{\otimes N}+\ket{1}^{\otimes N}\right)\right] \nonumber \\
        &=\frac{1}{2}[\bra{0}\mathcal{P}^{\prime\dagger}\otimes\bra{0}\mathcal{Q}^{\prime\dagger}_1\otimes\cdots\otimes\bra{0}\mathcal{Q}^{\prime\dagger}_{N-2}\otimes\bra{0}\nonumber\\
        &~~~~~~~~~~~~~~~~+\bra{1}\mathcal{P}^{\prime\dagger}\otimes\bra{1}\mathcal{Q}^{\prime\dagger}_1\otimes\cdots\otimes\bra{1}\mathcal{Q}^{\prime\dagger}_{N-2}\otimes\bra{1} ]. \nonumber \\
        &~~~~[\mathcal{P}^{\prime\prime}\ket{0}\otimes\mathcal{Q}^{\prime\prime}_1\ket{0}\otimes\cdots\otimes\mathcal{Q}^{\prime\prime}_{N-2}\ket{0}\otimes\ket{0}\nonumber\\
        &~~~~~~~~~~~~~~~~+\mathcal{P}^{\prime\prime}\ket{1}\otimes\mathcal{Q}^{\prime\prime}_1\ket{1}\otimes\cdots\otimes\mathcal{Q}^{\prime\prime}_{N-2}\ket{1}\otimes\ket{1}] \nonumber \\
        &=\frac{1}{2}[\bra{0}\mathcal{P}^{\prime\dagger}\mathcal{P}^{\prime\prime}\ket{0}\otimes\bra{0}\mathcal{Q}^{\prime\dagger}_1\mathcal{Q}^{\prime\prime}_1\ket{0}\otimes\dots\otimes\bra{0}\mathcal{Q}^{\prime\dagger}_{N-2} \nonumber \\
        &~~~~~~~~~\mathcal{Q}^{\prime\prime}_{N-2}\ket{0}
        +\bra{1}\mathcal{P}^{\prime\dagger}\mathcal{P}^{\prime\prime}\ket{1}\otimes\bra{1}\mathcal{Q}^{\prime\dagger}_1\mathcal{Q}^{\prime\prime}_1\ket{1}\nonumber \\
        &~~~~~~~~~~~~~~~~\otimes\dots\otimes\bra{1}\mathcal{Q}^{\prime\dagger}_{N-2}\mathcal{Q}^{\prime\prime}_{N-2}\ket{1}]
        \label{inner}
    \end{align}
    
    The following relations hold for Pauli operators:
    \begin{eqnarray}
        \expval{I}{0}=\expval{I}{1}=\expval{\sigma_Z}{0}=1; ~~ \expval{\sigma_Z}{1}=-1 \nonumber \\
        \expval{\sigma_X}{0}=\expval{\sigma_X}{1}=\expval{i\sigma_Y}{0}=\expval{i\sigma_Y}{1}=0~
        \label{relations}
    \end{eqnarray}
    When $\mathcal{M}=\mathcal{N}$, $\mathcal{P}^{\prime}=\mathcal{P}^{\prime\prime}$ and $\mathcal{Q}^{\prime}_k=\mathcal{Q}^{\prime\prime}_k~ \forall k$. This immediately leads to get $\mathcal{P}^{\prime\dagger}\mathcal{P}^{\prime\prime}=\mathcal{Q}^{\prime\dagger}_k\mathcal{Q}^{\prime\prime}_k=I ~ \forall k$. Therefore, using the relations of Eq. \ref{relations}, Eq. \ref{inner} becomes:
    \begin{align}
        &\left[\mathcal{M}\ket{\psi_N}\right]^{\dagger}\left[\mathcal{N}\ket{\psi_N}\right] \nonumber \\
        &=\frac{1}{2}[\expval{I}{0}\otimes\expval{I}{0}\otimes\cdots\otimes\expval{I}{0}\nonumber\\
        &~~~~~~~~~~~~~~~~~~+\expval{I}{1}\otimes\expval{I}{1}\otimes\cdots\otimes\expval{I}{1}] \nonumber \\
        &=\frac{1}{2}[1+1]=1
        \label{case0}
    \end{align}
    For the other condition $\mathcal{M}\neq\mathcal{N}$ we consider the following two cases as:\\
    \textbf{Case 1:} Suppose $\mathcal{P}^{\prime}=\mathcal{P}^{\prime\prime}$ then $\exists ~l \in [1,2,\cdots,N-2]$ s.t. $\mathcal{Q}^{\prime}_l\neq\mathcal{Q}^{\prime\prime}_l$. Hence, it follows that $\mathcal{Q}^{\prime\dagger}_l\mathcal{Q}^{\prime\prime}_l=\pm\sigma_X/i\sigma_Y$. Hence, Eq. \ref{inner} becomes:
    \begin{align}
        &\left[\mathcal{M}\ket{\psi_N}\right]^{\dagger}\left[\mathcal{N}\ket{\psi_N}\right] \nonumber \\
        &=\frac{1}{2}[\expval{I}{0}\otimes\cdots\expval{\pm\sigma_X/i\sigma_Y}{0}\otimes\cdots\expval{\mathcal{Q}^{\prime\dagger}_{N-2}\mathcal{Q}^{\prime\prime}_{N-2}}{0}\nonumber \\
        &~~~~+\expval{I}{1}\otimes\cdots\expval{\pm\sigma_X/i\sigma_Y}{1}\otimes\cdots\expval{\mathcal{Q}^{\prime\dagger}_{N-2}\mathcal{Q}^{\prime\prime}_{N-2}}{1}] \nonumber \\
        &=\frac{1}{2}[0+0]=0
        \label{case1}
    \end{align}
    
    \textbf{Case 2:} When $\mathcal{P}^{\prime}\neq\mathcal{P}^{\prime\prime}$ if $\exists ~l \in [1,2,\cdots,N-2]$ such that $\mathcal{Q}^{\prime}_l\neq\mathcal{Q}^{\prime\prime}_l$ then Eq.\ref{inner} identically vanishes just like Eq.\ref{case1}. If $\mathcal{Q}^{\prime}_k=\mathcal{Q}^{\prime\prime}_k~ \forall k$, then for $\mathcal{P}^{\prime\dagger}\mathcal{P}^{\prime\prime}=\pm\sigma_X/i\sigma_Y$ Eq. \ref{inner} again vanishes too. Now, when $\mathcal{P}^{\prime\dagger}\mathcal{P}^{\prime\prime}=\pm\sigma_Z$ then Eq. \ref{inner} becomes:
    
    \begin{align}
        &\left[\mathcal{M}\ket{\psi_N}\right]^{\dagger}\left[\mathcal{N}\ket{\psi_N}\right] \nonumber \\
        &=\frac{1}{2}[\expval{\pm\sigma_Z}{0}\otimes\expval{I}{0}\otimes\cdots\otimes\expval{I}{0}\nonumber\\
        &~~~~~~~~~~~~~~~~+\expval{\pm\sigma_Z}{1}\otimes\expval{I}{1}\otimes\cdots\otimes\expval{I}{1}] \nonumber \\
        &=\frac{1}{2}[1-1]=0
        \label{case2}
    \end{align}
    
    Hence, from Eqs. \ref{case0}, \ref{case1}, \ref{case2}, one finds that all the $2^N$ generated states are orthonormal to each other and they can be used for superdense coding of any N bit classical information. 
    
    Although the encoding of this classical information through Alice's operation need not have any unique technique, we provide an efficient scheme of encoding. First qubit operation $\mathcal{P}=I,\sigma_X,\sigma_Z,i\sigma_Y$ corresponds those information consisting $00,01,10,11$, respectively at leftmost 2 bits. For the remaining operations $\mathcal{Q}=I/\sigma_Z,\sigma_X/i\sigma_Y$ on i-th qubit will correspond $0,1$ bit respectively at (i+1)-th place of the total information string. For example, a five qubit information `$10110$' will be encoded by the unitary operator $\sigma_Z\otimes\sigma_X\otimes\sigma_X\otimes I$ or $\sigma_Z\otimes\sigma_X\otimes\sigma_X \otimes\sigma_Z$ or $\sigma_Z\otimes i\sigma_Y\otimes i\sigma_Y\otimes I$ or $\sigma_Z\otimes i\sigma_Y\otimes i\sigma_Y\otimes \sigma_Z$ as per the choice of Alice. 
    
    In the following, we show that this generalized protocol is optimal. For a N qubit system its dimension is $d=2^N$. Holevo bound for any such channel is $H=\log_2d=N$. Here, Alice possesses $N-1$ qubits while 1 qubit belongs to Bob. Therefore, the dimension of Alice's subsystem is $d_A=2^{N-1}$. Generalized GHZ state is a GME state having a marginal state $I/2$ for each of its qubits. This fact ensures that Alice's and Bob's subsystems are maximally entangled making the entropy of Bob's marginal maximum. Also, $\ket{\psi_N}$ being a pure state possesses zero entropy. Hence, from Eq. \ref{capacity} we get the capacity of the channel as $X=\log_22^{N-1}+S(\rho_B=\frac{I}{2})-S(\ket{\psi_N}\bra{\psi_N})=(N-1)+1-0=N$, which is exactly same as the Holevo bound.
    
\section{Security Analysis}\label{Security}
    \subsection{Different measurement basis for Alice and Bob}
    Interestingly, the two qubit Bell state ($\ket{\phi^+}$) and three qubit GHZ state can be written in $\{\ket{+},\ket{-}\}$ basis as:
    
    \begin{align}
        \ket{\psi_2}&=\frac{1}{\sqrt{2}}\left(\ket{00}+\ket{11}\right)=\frac{1}{\sqrt{2}}\left(\ket{++}+\ket{--}\right) \\
        \ket{\psi_3}&=\frac{1}{\sqrt{2}}\left(\ket{000}+\ket{111}\right)\nonumber\\
        &=\frac{1}{2}\left(\ket{+++}+\ket{+--}+\ket{-+-}+\ket{--+}\right)
    \end{align}
    One notices that all GHZ states written in $\{\ket{+},\ket{-}\}$ basis are superposition of all possible terms that contain even number of $\ket{-}$ states. In general N qubit GHZ state can be written as:
    \begin{align}
        \ket{\psi_N}&=\frac{1}{\sqrt{2}}\left(\ket{0}^{\otimes N}+\ket{1}^{\otimes N}\right)_{A_1\cdots A_{N-1}B} \label{Choice1} \\
        \ket{\psi_N}&=\frac{1}{\sqrt{2^{(N-1)}}}\sum_{a_1,\cdots ,a_N}^{1}\Big(\ket{a_1\cdots a_N}\nonumber\\
        &~~~~~~~~~~~~~~+(-1)^{\sum_{i=1}^N a_i}\ket{a_1\cdots a_N}\Big)_{A_1\cdots A_{N-1}B} \label{Choice2}
    \end{align}
    where, $\ket{a_i=0}=\ket{+}$ and $\ket{a_i=1}=\ket{-}$. Alice and Bob can utilize this different basis measurement technique for security analysis. First, Bob can measure in computational basis i.e. in $\{\ket{0},\ket{1}\}$ basis and we denote this choice as BMB$_1$ (Bob's measurement choice $1$). Correspondingly, Alice's computational basis choice will be denoted by AMB$_1$. On the other hand, if the state is measured as Eq. \ref{Choice2} then both AMB$_2$ and BMB$_2$ are $\{\ket{+},\ket{-}\}$. Using these two choices, Alice and Bob can construct a strategy to check for eavesdropping in the quantum channel.
    
    \subsection{Eavesdropping check}
    From a sufficiently large set of same states $\ket{\psi_N}$ distributed among Alice and Bob, Bob randomly chooses some of the states as a sample set to check whether Eve has anyhow altered the original state. He then measures his qubit either in BMB$_1$ or in BMB$_2$ and accordingly tells Alice about his measurement choice along with the corresponding outcome. For computational basis measurements (BMB$_1$ and AMB$_1$) Alice's outcome must collapse into $ \ket{0}^{\otimes N-1}$ or $\ket{1}^{\otimes N-1}$ according to his outcome $\ket{0}$ or $\ket{1}$. When Bob measures in BMB$_2$ and gets $\ket{+}$ then the outcome of AMB$_2$ must be a state which contains even number(s) of $\ket{-}$ and for $\ket{-}$ outcome of BMB$_2$ Alice must get odd number(s) of $\ket{-}$ as outcome of AMB$_2$. Finally, they check their corresponding outcomes and if for majority of the chosen states those remain consistent then they can securely use the remaining states for communication. Otherwise, the channel is corrupted and they need to abort further communication using them.\vspace{-0.4 cm}
    
    \subsection{Complete security check}
    Suppose, Eve still has some strategy to get the secret message sent by Alice. In this section, we show that Eve can not apply any other successful method to get the secret message without getting caught.
    
    In this protocol, Eve does not have access of all N qubits at the same time. After encoding the secret message, Alice physically sends her $N-1$ qubits to Bob. If Eve attacks those qubits, it will not be possible for her to extract the N bit secret message completely by doing any kind of operations or measurement on those $N-1$ qubits. So, her strategy will be to prepare some ancilla qubit $\ket{\xi}$ to intercept the N-th qubit which belongs to Bob. Then, when the $N-1$ qubits will be sent to Bob, she might capture those and make a joint measurement on those qubits along with her ancilla to get the secret information. Before Alice and Bob are engaged in the protocol let Eve operate some unitary operation U on Bob's qubit and her ancilla qubit $\ket{\xi}$. In most general form, we can write the operations as:
    \begin{align}
        U:&\ket{0}_B\ket{\xi}_E\longrightarrow\ket{0}_B\ket{\xi_{00}}_E+\ket{1}_B\ket{\xi_{01}}_E \nonumber \\
        &\ket{1}_B\ket{\xi}_E\longrightarrow\ket{0}_B\ket{\xi_{10}}_E+\ket{1}_B\ket{\xi_{11}}_E
        \label{unitary}
    \end{align}
    where, $\ket{\xi_{ij}}$ for $i,j\in{0,1}$ are ancilla states of Eve, completely determined by the unitary operation U.
    
    We now show that Eve fails to get any information for any N qubit GHZ state. After Eve's interaction with Bob's qubit, the generalized state takes the form:
    \begin{align}
        \ket{\psi_N^{\prime}}&=\frac{1}{\sqrt{2}}\big(\ket{0}^{\otimes N-1}(\ket{0}\ket{\xi_{00}}+\ket{1}\ket{\xi_{01}} \nonumber\\
        &~~~~~~~~+ \ket{1}^{\otimes N-1}\ket{0}\ket{\xi_{10}}+\ket{1}\ket{\xi_{11}}\big)_{A_1\cdots A_{N-1}BE} \nonumber \\
        &=\frac{1}{\sqrt{2}}\left(\ket{\eta_1}\ket{0}+\ket{\eta_2}\ket{1}\right)_{A_1\cdots A_{N-1}EB},
    \end{align}
    where,
    \begin{align}
        \ket{\eta_1}_{A_1\cdots A_{N-1}E}&=(\ket{0}^{\otimes N-1}\ket{\xi_{00}}+\ket{1}^{\otimes N-1}\ket{\xi_{01}}) \nonumber \\
        \ket{\eta_2}_{A_1\cdots A_{N-1}E}&=(\ket{0}^{\otimes N-1}\ket{\xi_{10}}+\ket{1}^{\otimes N-1}\ket{\xi_{11}})
        \label{eta}
    \end{align}
    As Bob measures in BMB$_1$ Alice's outcome for AMB$_1$ must be $\ket{0}^{\otimes N}$ or $\ket{1}^{\otimes N}$ according to Bob's outcome. The conditions for Eve to remain undiscovered while eavesdropping check by Alice and Bob must be,
    \begin{equation}
        \bra{1}^{\otimes N-1}\ket{\eta_1}=\bra{0}^{\otimes N-1}\ket{\eta_2}=0
        \label{condition3}
    \end{equation}
    Solving Eqs. \ref{eta} and \ref{condition3} we arrive at the following results,
    \begin{equation}
        \ket{\xi_{01}}=\ket{\xi_{10}}=\bm{0}.
        \label{result1}
    \end{equation}
   The generalized state after Eve's action can be expressed in AMB$_2$ and BMB$_2$ as:
    \begin{align}
        \ket{\psi_N^{\prime}}&= \frac{1}{\sqrt{2^{N+1}}}((\ket{+}+\ket{-})^{\otimes N}\ket{\xi_{00}}\nonumber\\
        &~~~~~~~~~~~~+ (\ket{+}-\ket{-})^{\otimes N}\ket{\xi_{11}})_{A_1\cdots A_{N-1}BE} \nonumber \\
        &=\frac{1}{\sqrt{2^{N+1}}}\left(\ket{\phi_1}\ket{+}+\ket{\phi_2}\ket{-}\right)_{A_1\cdots A_{N-1}EB}
    \end{align}
    where,
    \begin{align}
        \ket{\phi_1}&=\ket{\psi_{N-1}}\left(\ket{\xi_{00}}+\ket{\xi_{11}}\right)\nonumber\\
        &~~~~~~~~~~~~~~+\ket{\tilde{\psi}_{N-1}}\left(\ket{\xi_{00}}-\ket{\xi_{11}}\right)_{A_1\cdots A_{N-1}E} \nonumber \\
        \ket{\phi_2}&=\ket{\psi_{N-1}}\left(\ket{\xi_{00}}-\ket{\xi_{11}}\right)\nonumber\\
        &~~~~~~~~~~~~~~+\ket{\tilde{\psi}_{N-1}}\left(\ket{\xi_{00}}+\ket{\xi_{11}}\right)_{A_1\cdots A_{N-1}E}
        \label{phi}
    \end{align}
    Here, $\ket{\psi_N}$ is defined in Eq. \ref{Choice2} and $\ket{\tilde{\psi}_N}= \frac{1}{\sqrt{2^{(N-1)}}}\sum_{a_1,\cdots ,a_N}^{1}(\ket{a_1\cdots a_N}+(-1)^{\sum_{i=1}^N a_i+1} \ket{a_1\cdots a_N})_{A_1\cdots A_{N-1}B}$ with $\ket{a_i=0}=\ket{+}$ and $\ket{a_i=1}=\ket{-}$. Alice must obtain even (odd) number(s) of $\ket{-}$, when she measures $\ket{\phi_N}(\ket{\tilde{\phi}_N})$ in AMB$_2$. Hence, it must satisfy,
    \begin{align}
        \braket{p_1p_2\cdots p_{N-1}}{\phi_1}=\braket{q_1q_2\cdots q_{N-1}}{\phi_2}=0 \label{condition4} \nonumber \\
        s.t. \sum_{i=1}^{N-1} p_i=\text{even}, \sum_{i=1}^{N-1} q_i=\text{odd}
    \end{align}
    With the condition $\ket{p_i,q_i=0}=\ket{+} \& \ket{p_i,q_i=1}=\ket{-} \forall i$. From the construction it is evident that $\ket{\psi_N}(\ket{\tilde{\psi}_N})$ contains even (odd) number(s) of $\ket{-}$. Solving Eqs. \ref{phi} and \ref{condition4} we arrive at the conclusion that,
    \begin{equation}
        \ket{\xi_{00}}=\ket{\xi_{11}}.
        \label{result2}
    \end{equation}
    Applying the results of Eqs. \ref{result1} and \ref{result2} in Eq. \ref{unitary}, one obtains:
    \begin{align}
        U:&\ket{0}_B\ket{\xi}_E\longrightarrow\ket{0}_B\ket{\xi^{\prime}}_E \nonumber \\
        &\ket{1}_B\ket{\xi}_E\longrightarrow\ket{1}_B\ket{\xi^{\prime}}_E \nonumber
    \end{align}
    This clearly shows that the joint state of Bob and Eve will be a product state, as U is effectively a local unitary operation for Eve. As a result, we can deduce that this protocol is absolutely secure and effective under the assumption that the channel is not noisy and no information is getting lost within it.
    
\section{Distributed superdense coding technique}\label{D}
    As we compare the two dense coding protocols described in Section \ref{N-separable} (Protocol 1) and \ref{GME} (Protocol 2), we notice that absolutely secure generalized superdence coding is possible using two complete different types of states, one N-separable and another GME state. The only common feature they both possess is that both of these states show maximal entanglement across the bipartition of Alice and Bob. Noticeably, protocol 1 is only applicable to transfer even bits of information whereas protocol 2 is applicable for any N. In terms of the number of qubits needed to physically transfer from Alice to Bob, protocol 1 using N-separable states requires only $N/2$ (N even) qubits which is much less (for $N>2$) compared to $N-1$ qubits for the other one. However, Protocol 2 using GHZ states is more beneficial in different type of scenarios. Suppose, the original N bit classical information is distributed among $N-1$ spatially separated parties and all of them wants to send it to Bob.
    
    In this scenario, the proposed protocol 2 using generalized GHZ state will be the most efficient one. Since, all the unitaries operated to perform this protocol act only on individual qubits and no global multiqubit operations are required, therefore, all those $N-1$ qubits which are supposed to be with Alice, can now be distributed among those spacially separated $N-1$ sender parties. Now, each of them would apply the operation according to the protocol and send their individual qubit to Bob. Only one party who has 2 bits of information (as N bit information is distributed among $N-1$ parties) is able to operate all of those 4 operation choices mentioned in Section \ref{GME}. In order to send the information to Bob, other than using this protocol they can either create separate quantum channels between Bob and each one of them or they might send all their information to one chosen sender and she can transmit it to Bob. But in both cases total required qubits will be much greater than N. 
    
    Immediate next question that arises is whether protocol 1 is applicable in this type of scenario. Noticing the structure of the state $\ket{\psi_{2N}}$ used in Section \ref{N-separable}, increasing just one more qubit to Alice's part would result in Alice containing one complete EPR pair ($\ket{\phi^+}$). That would make the marginal state of Alice of the form Eq. \ref{separable} and the protocol will not remain absolutely secure anymore. Therefore, protocol 1 must be carried out only by keeping $\frac{N}{2}$ to both the parties.
    
    The number of parties among whom the initial information can be distributed may vary from 1 to $N-1$. We provide a distributed superdense coding method optimizing the number of qubits transferred and number of senders by combining protocol 1 and 2. We define this distributed generalized dense coding scheme D(N,k) where N is the size of information bit and $1\leq k\leq N-1$ is the number of parties between whom the information is distributed. The quantum state required for this scheme is given by,
    \begin{align}
        \ket{\psi_{D(N,k)}}=&\frac{1}{\sqrt{2}}\left(\ket{0}^{\otimes i}+\ket{1}^{\otimes i}\right)_{A_1\cdots A_{i-1}B}\otimes \ket{\phi^+}^{\otimes \frac{N-i}{2}}_{AB};\nonumber\\ &i=\max(2,2(k+1)-N) ~~ \forall~\text{N = even} \nonumber \\
        \ket{\psi_{D(N,k)}}=&\frac{1}{\sqrt{2}}\left(\ket{0}^{\otimes j}+\ket{1}^{\otimes j}\right)_{A_1\cdots A_{j-1}B}\otimes \ket{\phi^+}^{\otimes \frac{N-j}{2}}_{AB};\nonumber \\ &j=\max(3,2(k+1)-N) ~~ \forall~\text{N = odd}
    \end{align}
    When $k\leq\frac{N}{2}$, for even N $\ket{\psi_{D(N,k)}}$ becomes exactly similar as $\ket{\psi_{2N}}$ of protocol 1 with $\frac{N}{2}$ EPR pairs. For odd N it is a product state of one 3 qubit GHZ state and $\frac{N-3}{2}$ pairs of $\ket{\phi^+}$. When $k>\frac{N}{2}$, $\ket{\psi_{D(N,k)}}$ becomes a $(N-k)$-separable state containing $N-(k+1)$ pairs of $\ket{\phi^+}$ and a GHZ state of $2(k+1)-N$ qubits. All encoding technique for GHZ states will be like protocol 2 where, that for EPR pairs will follow protocol 1. For $k\leq\frac{N}{2}$ number of qubits belonging to Alice is always $\lceil\frac{N}{2}\rceil$, hence those qubits can be distributed among k parties according to their information content. For $k>\frac{N}{2}$ total number of qubits Bob possess is $N-k$ ($1$ from GHZ part and $N-k-1$ from EPR part). Therefore, the remaining k qubits will be equally distributed among k separated parties. If we notice Bob's marginal state for $\ket{\psi_{D(N,k)}}$ it will be $\rho_B=\frac{I}{2}\otimes \frac{I}{2}^{\otimes \left[\frac{N-2}{2}\right]}={I_{2^{\left[\frac{N}{2}\right]}}}/{2^{\left[\frac{N}{2}\right]}}$ for $k\leq\frac{N}{2}$ and $\rho_B=\frac{I}{2}\otimes \frac{I}{2}^{\otimes N-(k+1)}=\frac{I_{N-k}}{2^{N-k}}$ for $k>\frac{N}{2}$. It ensures the maximal entanglement condition across Alice's and Bob's bipartition, making D(N,k) optimal.\vspace{0.5 cm}
    
\section{Conclusion}
    In conclusion, we have shown that for any superdense coding protocol using two qubit systems, maximal entanglement is necessary i.e., the state must be an AME state. For a general N qubit system $(N>2)$ we established that Bob's subsystem with dimension d must have the maximally mixed marginal $I_d/d$, making their bipartition maximally entangled. This is the only condition required for optimal dense coding to happen. To show that neither AME nor GME conditions are required we have briefly described a protocol to send even bit of information using an N-separable state. We then presented a new dense coding technique to send a general N bit classical information using genuinely entangled GHZ states. Later we have shown that this scheme is completely secure from eavesdropper's attack. 
    
    Furthermore, considering a different type of scenario where the initial information is split among different spatially separated parties, our method using GHZ state is the only applicable dense coding method known for these scenarios. We further generalized this method to provide the distributed superdense coding protocol D(N,k) to optimize number of qubits sent to Bob and the number of parties the information is distributed among. Hence, this work provides a complete idea of absolutely secure one way quantum communication between two or more spatially separated parties through Distributed superdence coding.
    
    \section{Acknowledgements}
    The authors acknowledge Mr. Abhinash Kumar Roy and Mr. Soumik Mahanty of IISER Kolkata for many encouraging and insightful discussions. They would also like to express their gratitude for the financial support from DST, India through Grant No. DST/ICPS/QuST/Theme1/2019/2020-21/01 and to the hospitality of IISER Kolkata.
    
    \bibliography{sample}
    
\end{document}